\documentclass[aps,preprint]{revtex4}
\usepackage[dvips]{graphicx}
\usepackage{latexsym}
\usepackage{natbib}
\usepackage[centerlast]{caption}
\usepackage{subfig}

\newcommand{\Rmnum}[1]{\expandafter\@slowromancap\romannumeral#1@}

\begin{document}

\title{Robustness of interdependent networks under targeted attack}

\author{Xuqing Huang,$^1$ Jianxi Gao,$^{1,2}$ Sergey V. Buldyrev,$^3$ Shlomo Havlin,$^4$ and
H.Eugene Stanley$^1$}

\affiliation{$^1$Center for Polymer Studies and Department of Physics, Boston University, Boston, MA 02215 USA\\
$^2$Department of Automation, Shanghai Jiao Tong University, 800 Dongchuan Road, Shanghai, 200240, PR China\\
$^3$Department of Physics,~Yeshiva University, New York, NY 10033 USA\\
$^4$Minerva Center and Department of Physics, Bar-Ilan University, 52900 Ramat-Gan, Israel}


\begin{abstract}
When an initial failure of nodes occurs in interdependent networks, a cascade of failure between the networks occurs. 
Earlier studies focused on random initial failures. Here we study the robustness of interdependent networks under targeted attack on high or low degree nodes. We introduce a general technique and show that the {\it targeted-attack} problem in interdependent networks can be mapped to the {\it random-attack} problem in a transformed pair of interdependent networks.
We find that when the highly connected nodes are protected and have lower probability to fail, in contrast to single scale free (SF) networks where the percolation threshold $p_c=0$, coupled SF networks are significantly more vulnerable with $p_c$ significantly larger than zero. The result implies that interdependent networks are difficult to defend by strategies such as protecting the high degree nodes that have been found useful to significantly improve robustness of single networks.
\end{abstract}
\maketitle
Modern systems due to  technological progress are becoming more and more mutually coupled and depend on each other to provide proper functionality~\cite{Laprie2007,Panzieri2008,Rosato2008}. Social disruptions caused by recent disasters, ranging from hurricanes to large-scale power outages and terrorist attacks, have shown that the most dangerous vulnerability is hiding in the many interdependencies across different networks~\cite{Chang2009}.
The question of robustness of interdependent networks has recently become of interest~\cite{Leicht2009,Vespignani2010,Sergey2010,Rony2010}.
In interdependent networks, nodes from one network depend on nodes from another network and vice versa. Consequently, when nodes from one network fail they cause nodes in the other network to fail too. When some initial failure of nodes happens, this may trigger a recursive process of cascading failures that can completely fragment both networks.

Recently, a theoretical framework was developed~\cite{Sergey2010} to study the process of cascading failures in interdependent network caused by {\it random} initial failure of nodes. They show that due to the coupling between networks, interdependent networks are extremely vulnerable to random failure.
However, when we consider real scenarios, initial failure is mostly not random. It may be due to a {\it targeted attack} on important central nodes. It can also occur to low central nodes because important central nodes are purposely defended, e.g. in internet networks, heavily connected hubs are purposely more secured. Indeed, it was shown that targeted attacks on high degree nodes~\cite{Barabasi2000,Callaway2000,Cohen2001,Gallos2005,Moreira2009} or high betweeness nodes~\cite{Holme2002} in {\it single} networks have dramatic effect on their robustness. The question of robustness of {\it interdependent} networks under targeted attack or defense has not been addressed.

In this Letter, we develop a mathematical framework for understanding the robustness of interdependent networks under initial targeted attack which depends on degree of nodes. The framework is  based on a general technique we develop to solve targeted attack problems in networks by mapping them to random attack problems.
A value $W_{\alpha}(k_i)$ is assigned to each node, which represents the probability that a node $i$ with $k_i$ links is initially attacked and become inactive. We focus on the family of functions~\cite{Gallos2005}
\begin{equation}
  W_{\alpha}(k_i)=\frac{k_i^\alpha}{\sum_{i=1}^{N}k_i^\alpha}, -\infty<\alpha<+\infty .
  \label{Eq1}
\end{equation}
When $\alpha>0$, nodes with higher degree are more vulnerable and those nodes are intentionally attacked, while for $\alpha<0$, nodes with higher degree are defended and so have lower probability to fail. The case $\alpha=0$, $W_0=\frac{1}{N}$, represents the random removal of nodes ~\cite{Sergey2010} and the case $\alpha\to\infty$ represents the targeted attack case where nodes are removed strictly in the order from high degree to low degree. For the $\alpha<0$ case, nodes with zero degree should be removed before analysis begins.

Our model consists of two networks, A and B, with the same number of nodes, $N$. The $N$ nodes in each network are connected to nodes in the other network by bidirectional dependency links, thereby establishing a one-to-one correspondence. The functioning of a A-node in network A depends on the functioning of the corresponding B-node in network B and vice versa.
Within each network, the nodes are randomly connected with degree distributions $P_A(k)$ and $P_B(k)$ respectively. 
We begin by studying the situation where only network A is attacked. We initially remove a fraction, $1-p$, of the A-nodes of network A with probability $W_{\alpha}(k_i)$ (Eq.(\ref{Eq1})) and remove all the A-links that connect to those removed nodes. As nodes and links are sequentially removed, network A begins to fragment into connected components. Nodes that are not connected to the giant component are considered inactive and are removed. Owing to the dependence between the networks, all the B-nodes in network B that are connected to the removed A-nodes in network A are then also removed. Network B also begins to fragment into connected components and only the nodes in the giant component are kept. Then network B spreads damage back to network A. The damage is spreaded between network A and B, back and forth until they completely fragment or arrive to a mutually connected component and no further removal of nodes and links occurs.

The main idea of our approach is to find an equivalent network $A'$, such that the {\it targeted} attack problem on interdependent networks $A$ and $B$ can be solved as a {\it random} attack problem on interdependent networks $A'$ and $B$.
We start by finding the new degree distribution of network $A$ after removing, according to Eq.(\ref{Eq1}), $1-p$ fraction of nodes but before the links of the remaining nodes which connect to the removed nodes are removed. Let $A_p(k)$ be the number of nodes with degree $k$ and $P_p(k)$ be the new degree distribution of the remaining fraction $p$ of nodes in network $A$,
\begin{equation}
  P_p(k)=\frac{A_p(k)}{pN}.
  \label{Eq2}
\end{equation}
When another node is removed, $A_p(k)$ changes as
\begin{equation}
  A_{(p-1/N)}(k)=A_p(k)-\frac{P_p(k)k^\alpha}{<k(p)^\alpha>},
  \label{Eq3}
\end{equation}
where $<k(p)^\alpha> \equiv \sum P_p(k)k^\alpha$. In the limit of $N\to\infty$, Eq.(\ref{Eq3}) can be presented
in terms of derivative of $A_p(k)$ with respect to $p$,
\begin{equation}
  \frac{dA_p(k)}{dp}=N\frac{P_p(k)k^\alpha}{<k(p)^\alpha>}.
  \label{Eq4}
\end{equation}
Differentiating Eq.(\ref{Eq2}) with respect to $p$ and using Eq.(\ref{Eq4}), we obtain
\begin{equation}
  -p\frac{dP_p(k)}{dp}=P_p(k)-\frac{P_p(k)k^{\alpha}}{<k(p)^\alpha>},
  \label{Eq5}
\end{equation}
which is exact for $N\to\infty$. In order to solve Eq.(\ref{Eq5}), we define a function $G_{\alpha}(x) \equiv \sum_k P(k)x^{k^\alpha}$, and substitue $f\equiv G_\alpha^{-1}(p)$. We find by direct differentiation that~\cite{Shao2009}
\begin{equation}
  P_p(k)=P(k)\frac{f^{k^\alpha}}{G_\alpha(f)}
  =\frac{1}{p}P(k)f^{k^\alpha},
  \label{Eq6}
\end{equation}
\begin{equation}
  <k(p)^\alpha>=\frac{fG_\alpha'(f)}{G_\alpha(f)},
\end{equation}
satisfy the Eq.(\ref{Eq5}). With this degree distribution, the generating function of the nodes left in network $A$ before removing the links to the removed nodes is
\begin{equation}
  G_{Ab}(x) \equiv \sum_kP_p(k)x^k=\frac{1}{p}\sum_kP(k)f^{k^\alpha}x^k.
  \label{Eq7}
\end{equation}
On the other hand, the fraction of the orginal links that connect to the nodes left is
\begin{equation}
  \tilde{p} \equiv \frac{pN<k(p)>}{N<k>}=\frac{\sum_kP(k)kf^{k^\alpha}}{\sum_kP(k)k},
  \label{Eq8}
\end{equation}
where $<k>$ is the average degree of the original network $A$, $<k(p)>$ is the average degree of remaining nodes before the links that are disconnected are removed.
So the generating function of the new degree distribution of the nodes left in network $A$ after their links to the removed nodes are also removed is~\cite{Newman2002} 
\begin{equation}
  G_{Ac}(x) \equiv G_{Ab}(1-\tilde{p}+\tilde{p}x).
  \label{Eq9}
\end{equation}
The only difference in the cascading process under {\it targeted} attack from the case under {\it random} attack is the first stage where the initial attack is exerted on the network $A$. If we find a network $A'$ with generating function $\tilde{G}_{A0}(x)$, such that after a random attack with $(1-p)$ fraction of removed, the generating function of nodes left in $A'$ is the same as $G_{Ac}(x)$, then the targeted attack problem on interdependent networks $A$ and $B$ can be solved as a random attack problem on interdependent networks $A'$ and $B$. We find $\tilde{G}_{A0}(x)$ by solving the equation $\tilde{G}_{A0}(1-p+px)=G_{Ac}(x)$ and from Eq.(\ref{Eq9}),
\begin{equation}
  \tilde{G}_{A0}(x)=G_{Ab}(1+\frac{\tilde{p}}{p}(x-1)).
  \label{Eq10}
\end{equation}
Up to now, we have mapped the problem of cascade of failures of nodes in interdependent networks caused by initial {\it targeted} attack to the problem of {\it random} attack.  We can see that the evolution of equations only depends on the generating function of network $A$, but not on any information about how the two networks interact with each other. Thus, this approach can be generally applied to study both single networks and other interdependent network models.

Next we can apply the framework developed in Ref.~\cite{Sergey2010},
$g_A(p)=1-\tilde{G}_{A0}[1-p(1-f_A)]$, where $f_A$ is a function of $p$ that satisfies the transcendental equation $f_A=\tilde{G}_{A1}[1-p(1-f_A)]$. Analogous equations exist for network B.
As the interdependent networks achieve a mutually connected giant component, the fraction of nodes left in giant component is $p_\infty$. The system satisfies the equations
\begin{equation}
\begin{array}{l}
x=pg_A(y),\\
y=pg_B(x),
\end{array}
\label{Eq11}
\end{equation}
where the two unknown variables $x$ and $y$ satisfy $p_\infty=xg_B(x)=yg_A(y)$. Eliminating $y$ from
these equations, we obtain a single equation
\begin{equation}
  x=pg_A[pg_B(x)].
  \label{Eq12}
\end{equation}
The critical case ($p=p_c$) emerges when both sides of this equation have equal derivatives,
\begin{equation}
  1=p^2\frac{dg_A}{dx}[pg_B(x)]\frac{dg_B}{dx}(x)|_{x=x_c,p=p_c}.
  \label{Eq13}
\end{equation}
which, together with Eq.(\ref{Eq12}), yields the solution for $p_c$ and the critical size of the giant mutually connected component, $p_\infty(p_c)=x_cg_B(x_c)$. In general, there is no explicit expression as a solution and $p_c$ and $x_c$ can be found numerically.

\begin{figure}[h!]
\centering \includegraphics[width=1\textwidth]{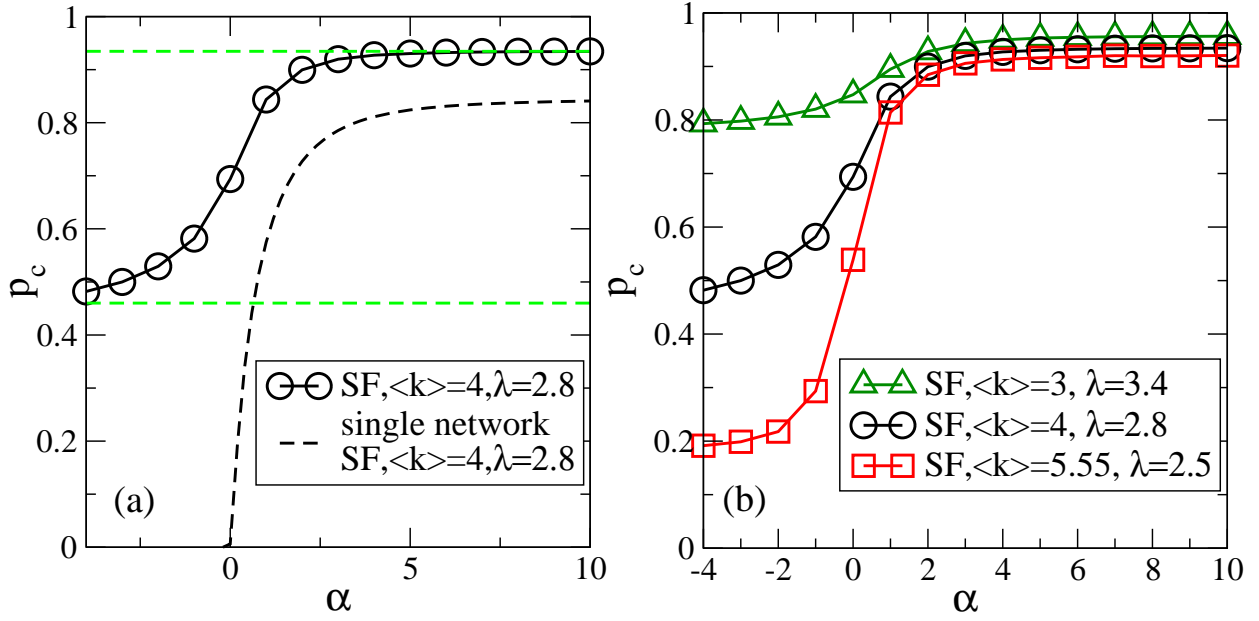}
\caption{(a) Dependence of $p_c$ on $\alpha$ for SF single and interdependent networks with average degree $<k>=4$. The lower cut-off of the degree is $m=2$.
The horizontal lines represent the upper and lower limits of $p_c$. The black dashed line represents $p_c$ for SF free network.
(b) Values of $p_c$ vs $\alpha$ for SF interdependent networks with different $\lambda$ and lower cut-off $m=2$. The $\lambda$ in the legends of both the graphs are approximate numbers.}
\label{figs1}
\end{figure}

We now analyze the specific classes of Erd\H{o}s-R\'{e}nyi (ER)~\cite{ER1959,Bollobas1985} and scale-free (SF)~\cite{SF1999,Caldarelli2007,Havlin2010} networks.
The lines in Fig.~\ref{figs1} represent the critical thresholds, $p_c$, for coupled coupled SF networks with different $\alpha$ obtained by solutions of Eq.(\ref{Eq12}) and Eq.(\ref{Eq13}), which are in excellent agreement with simulations. Several conclusions from Fig.~\ref{figs1} are as follows: (i) Remarkably, while $p_c$ for a single SF network approaches to $0$ quickly when $\alpha$ becomes zero or negative (see also \cite{Gallos2005}), $p_c$ for interdependent networks is non-zero for the entire range of $\alpha$ (Fig.~\ref{figs1}(a)). This follows from the fact that failure of the least connected nodes in one network may lead to failure of well connected nodes in the other network, which makes interdependent networks significantly more difficult to protect compared to a single network. 
(ii) targeted attacks ($\alpha>0$) and defense strategies ($\alpha<0$) are more effective for interdependent networks with broader degree distributions. In  Fig.~\ref{figs1}(b), comparing the lines of $\lambda=2.5$, $\lambda=2.8$ and $\lambda=3.4$ with $m=2$, one can see that the lower is $\lambda$ the more sensitive is $p_c$ to the change of $\alpha$. Accordingly, robustness of interdependent networks with broader degree distributions decreases more under the same targeted attacks.

Simplified forms for $G_{Ab}(x), G_{Ac}(x)$ and $\tilde{G}_{A0}(x)$ from Eqs.(\ref{Eq7}),(\ref{Eq9}) and (\ref{Eq10}) exist when $\alpha=1$,
\begin{equation}
	G_{Ab}(x)=\frac{1}{p}\sum_kP(k)f^kx^k=\frac{1}{p}G_{A0}(fx),\\
	\label{Eq14}
\end{equation}
\begin{equation}
	G_{Ac}(x)=\frac{1}{p}G_{A0}(f(1-\tilde{p}+\tilde{p}x)),\\
  	\label{Eq15}
\end{equation}
\begin{equation}
  	\tilde{G}_{A0}(x)=\frac{1}{p}G_{A0}(\frac{\tilde{p}}{p}f(x-1)+f).
  \label{Eq16}
\end{equation}
where $G_{A0}(x)$ is the original generating function of the network A, $f=G_{A0}^{-1}(p)$ and $\tilde{p}=\frac{G_{A0}'(f)}{G_{A0}'(1)}f$.

Explicit solutions of percolation quantities exist for the case of interdependent Erd\H{o}s-R\'{e}nyi networks, when $\alpha=1$ and both of the two networks are initially attacked simutaneously. The two networks originally have generating functions $G_{A0}(x)$ and $G_{B0}(x)$.
Initially, $(1-p_1)$ and $(1-p_2)$ fraction of nodes are targeted (according to Eq. (1) and $\alpha=1$) and removed from network A and B respectively. Similarly, we start by finding the equivalent networks $A'$ and $B'$ such that a fraction $(1-p_1p_2)$ of random initial attack on both of the networks has the same effect as $(1-p_1)$ and $(1-p_2)$ fraction of nodes are intentionally removed from network A and network B respectively.
After removal of initially failed nodes and all the links that connect to the removed nodes, according to Eq.(\ref{Eq15}), the generating function of the nodes left in network $A$ is
\begin{equation}
	G_{Ac}(x)=\frac{1}{p_1}G_{A0}(f_1(1-\tilde{p}_1+\tilde{p}_1x)),
	\label{Eq17}
\end{equation}
where $f_1 \equiv G_{A0}^{-1}(p_1)$, $\tilde{p}_1 \equiv f_1\frac{G_{A0}'(f_1)}{G_{A0}'(1)}$. 
Furthermore, $(1-p_2)$ fraction of the remaining A-nodes are randomly removed. Because each remaining A-node's corresponding B-node in network B has a possibility $(1-p_2)$ to be initially attacked, which leads to fail this A-node. The generating function of the nodes left in network $A$ is
\begin{equation}	
	G_{Ad}(x) \equiv G_{Ac}(1-p_2+p_2x)=\frac{1}{p_1}G_{A0}(f_1+\tilde{p}_1f_1p_2(x-1)).
	\label{Eq18}
\end{equation}
Now we can find the generating function of the equivalent network $A'$ by $\tilde{G}_{A0}(1-p_1p_2+p_1p_2x)=G_{Ad}(x)$:
\begin{equation}	
	\tilde{G}_{A0}(x)=\frac{1}{p_1}G_{A0}(\frac{\tilde{p_1}}{p_1}f_1(x-1)+f_1).
	\label{Eq19}
\end{equation}
The same holds for network $B'$.

For ER networks, the generating function is $G_0(x)=e^{<k>(x-1)}$~\cite{Newman2002},
so $f_1=\frac{ln(p_1)}{<k>_1}+1$,$f_2=\frac{ln(p_2)}{<k>_2}+1$, $\tilde{G}_{A0}(x)=\tilde{G}_{A1}(x)=e^{<k>_1f_1^2(x-1)}$ and $\tilde{G}_{B0}(x)=\tilde{G}_{B1}(x)=e^{<k>_2f_2^2(x-1)}$. From Eq.(\ref{Eq11}),
\begin{equation}
\begin{array}{l}
x=p_1p_2g_A(y)=p_1p_2(1-f_A),\\
y=p_1p_2g_B(x)=p_1p_2(1-f_B),
\end{array}
\label{Eq20}
\end{equation}
where
\begin{equation}
\begin{array}{l}
f_A=e^{<k>_1f_1^2y(f_A-1)},\\
f_B=e^{<k>_2f_2^2x(f_B-1)}.
\end{array}
\label{Eq21}
\end{equation}

\begin{figure}[h!]
\centering \includegraphics[width=1\textwidth]{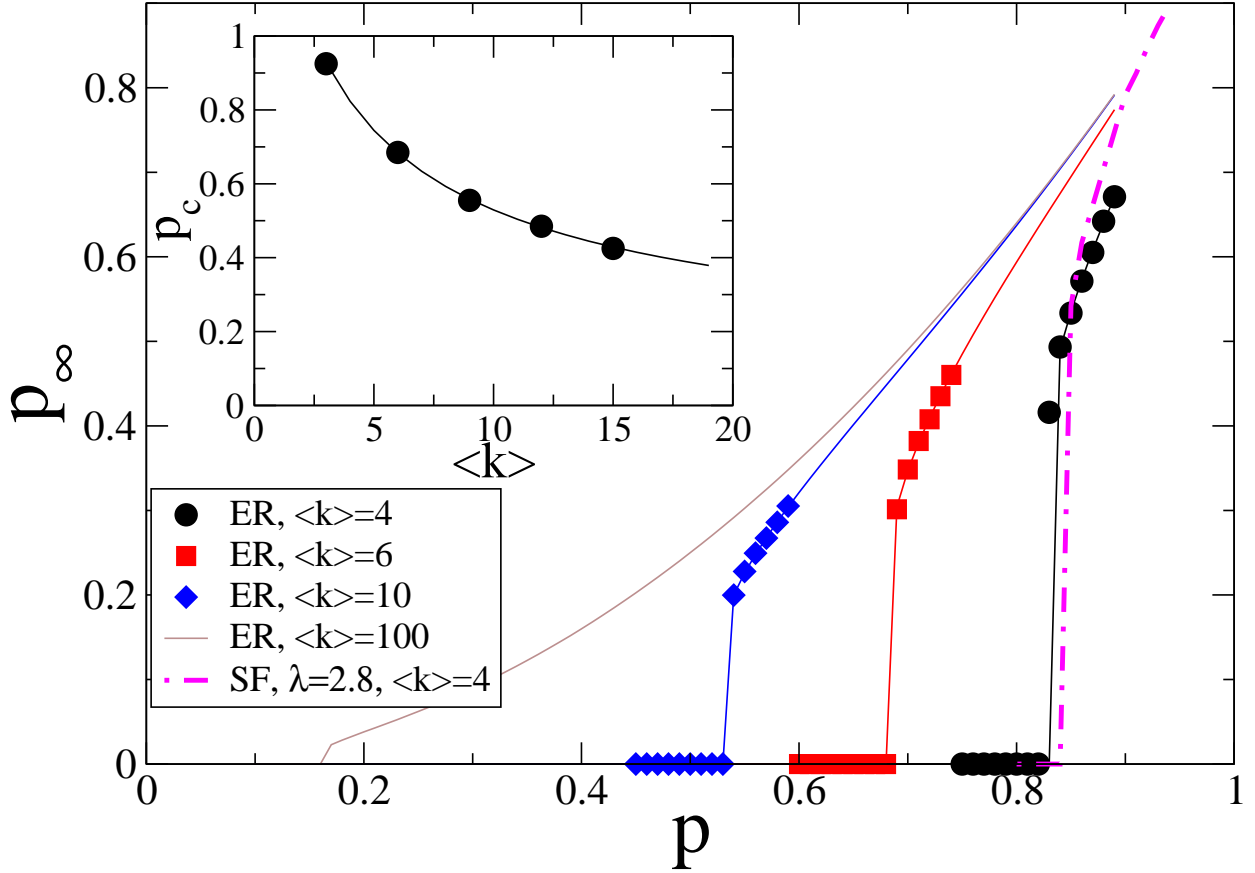}
\caption{Values of $p_{\infty}$ vs $p$ when both networks are initially attacked. Both networks in the interdependent networks are ER or SF networks with the same average degree. The symbols represent simulation data ($N=10^6$ nodes). 
The solid lines are theoretical predictions, Eq.(\ref{Eq22}). The dashed line represents simulation data for interdependent scale-free networks with $\lambda=2.8$, $<k>=4$. All results are for $\alpha=1$. Inset: Values of $p_c$ {\it vs} average degree of ER networks. The symbols represent simulation data, while the line is the theory, Eq.(\ref{Eq23}). }
\label{figs3}
\end{figure}

In the case $<k>_1=<k>_2=<k>$ and $p_1=p_2=p$, we find that
\begin{equation}	
	p_\infty=p^2(1-e^{<k>f^2p_\infty}).
	\label{Eq22}
\end{equation}
where $f_1=f_2 \equiv f=\frac{ln(p)}{<k>}+1$, and $p_c$ satisfies relation:
\begin{equation}	
	<k>p_c^2f_c=2.4554,
	\label{Eq23}
\end{equation}
with $f_c=\frac{ln(p_c)}{<k>}+1$.
Fig.~\ref{figs3} shows that the simulation confirms well the theory.
Compared to the case of random attack on one network, where $p_c=2.4554/<k>$~\cite{Sergey2010}, in Eq.(\ref{Eq23}), the factor $f_c$ reflects the effect of targeted attack on high degree nodes to increase $p_c$. The term $p_c^2$ in Eq.(\ref{Eq23}) is since we are initially attacking both networks simutaneously instead of only attacking one network. Indeed for the case of initial random attack on two networks simultaneously, from Eq.(\ref{Eq20}) and $f_A=e^{<k>_1y(f_A-1)}$, $f_B=e^{<k>_1y(f_B-1)}$~\cite{Sergey2010} we obtain $<k>p_c^2=2.4554$.

In summary, we developed a theoretical framework for understanding the robusteness of interdependent networks under targeted attacks on specific degree nodes.
We introduce a method and show that targeted-attack problems in networks can be mapped to random-attack problems by transforming the networks which are under initial attack. It provides a routine method (if the random-attack case is solvable) to study the targeted-attack problems in both single networks and randomly connected and uncorrelated interdependent networks, i.e. (i) the case of three or more interdependent networks, (ii) the case of partially coupled interdependent networks, (iii) the case in which a node from network $A$ can depend on more than one node from network $B$.
By applying the method, we find that in contrast to single networks, when the highly connected nodes are defended ($\alpha<0$), the percolation threshold $p_c$ has a finite non-zero value which is significantly larger than zero. For example, when the degrees of all nodes are known and nodes can only be damaged from lower degree to high degree ($\alpha\rightarrow-\infty$), $p_c\approx0.46$ for coupled SF networks with $\lambda=2.8$ and $<k>=4$ while $p_c$ for the same single SF network is $0$ (Fig.~\ref{figs1}). The implications of the present study are dramatic. The current methods applied to design robust networks and improve the robustness of current networks, i.e. protecting the high degree nodes, need to be modified to apply to interdependent network systems.


\newpage

\end{document}